\documentclass[prod]{jfmfr}
\usepackage{epsfig}
\usepackage[english]{babel}
\usepackage{textcomp}
\usepackage{amsmath}
\usepackage{color}
\usepackage{amssymb}
\usepackage{hyperref}
\usepackage{epic}
\usepackage{pict2e}
\usepackage{epstopdf}
\usepackage{pict2e} 
\usepackage{ulem}

\def\v{\mathbf{v}}
\def\vdp{\mathbf{v}_{\mathrm{dp}}}
\def\tmix{T_\text{mix}}
\def\peff{Pe_\mathit{eff}}
\def\deff{D_\mathit{eff}}

\usepackage{graphics}
\usepackage{amsfonts}
\usepackage{amssymb}
\usepackage{fancybox}
\usepackage{pstricks}
\usepackage{rotating}
\usepackage{tikz}

\begin{document}


\shorttitle{Diffusiophoresis, Batchelor scale and effective P\'eclet numbers}
\shortauthor{Florence Raynal and Romain Volk}

\title{Diffusiophoresis, Batchelor scale and effective P\'eclet numbers}

\author{Florence Raynal\aff{1} \corresp{\email{florence.raynal@ec-lyon.fr}}
 \and Romain Volk\aff{2} \corresp{\email{romain.volk@ens-lyon.fr}}
}
 
\affiliation{\aff{1}LMFA, Univ Lyon, \'Ecole Centrale Lyon, INSA Lyon, Universit\'e Lyon 1, CNRS,\\ F-69134 \'Ecully, France. \aff{2}Laboratoire de Physique, ENS de Lyon, Univ Lyon, CNRS, 69364 Lyon CEDEX 07, France.}

\date{4 April 2019; revised 16 July 2019; accepted 17 July 2019} 
\pubyear{2019} 
\volume{876}
\pagerange{818--829}
\doi{jfm.2019.589}

\maketitle

\begin{abstract}
We study the joint mixing of colloids and salt released together in a stagnation point or in a globally chaotic flow. 
In the presence of salt inhomogeneities, the mixing time is strongly modified depending on the sign of the diffusiophoretic coefficient $D_\mathrm{dp}$. Mixing is delayed when $D_\mathrm{dp}>0$ (salt-attracting configuration), or faster when $D_\mathrm{dp}<0$ (salt-repelling configuration). In both configurations, as for molecular diffusion alone, large scales are barely affected in the dilating direction while the Batchelor scale for the colloids, $\ell_{c,\mathrm{diff}}$, is strongly modified by diffusiophoresis. 
We propose here to measure a global effect of diffusiophoresis in the mixing process through an effective P\'eclet number built on this modified Batchelor scale.
Whilst this small scale is obtained analytically for the stagnation point, in the case of chaotic advection, we derive it using the equation of gradients of concentration, following Raynal \& Gence (\textit{Intl J. Heat Mass Transfer}, vol. 40 (14), 1997, pp. 3267--3273). 
Comparing to numerical simulations, we show that the mixing time can be predicted by using the same function as in absence of salt, but as a function of the effective P\'eclet numbers computed for each configuration. The approach is shown to be valid when the ratio $D_\mathrm{dp}^2/D_s D_c \gg 1$, where $D_c$ and $D_s$ are the diffusivities of the colloids and salt.

\end{abstract}


\section{Introduction}
{Mixing is the operation by which inhomogeneities of a scalar are attenuated by the combined action of advection by a flow and molecular diffusion. In this context, it is interesting to predict the time needed to achieve mixing as a function of the P\'eclet number, $Pe$, which has been done for a variety of simple flows at the heart of our understanding of mixing, and for chaotic advection by laminar flows ((\cite{bib:Metcalfeetal2012,bib:sundararajan2012,bib:villermaux2019})).}

In the case of mixing of colloids in the presence of salt inhomogeneities, the situation is more complex as salt gradients can enhance or delay mixing depending on whether colloids and salt are released in the same patch or with complementary profiles (\cite{bib:Abecassisetal2009,bib:Deseigneetal2014,bib:Volketal2014,bib:Maugeretal2016}). 
This is a situation which occurs frequently in microfluidic devices where salt and buffer are added to stabilize colloidal suspensions. 
Although this strong modification of the mixing time is due to compressible effects, which act as a source of scalar variance (\cite{bib:Volketal2014}), attempts were made to describe diffusiophoresis as an effective diffusion, and \cite{bib:Deseigneetal2014} proposed to rescale the mixing time by using an effective P\'eclet number for the colloids based on the expression of the effective diffusivity derived in \cite{bib:Abecassisetal2009}. Using the Ranz model of mixing (\cite{bib:ranz1979}), they were able to achieve a good rescaling of the mixing time measured in the salt-repelling case, but no expression was proposed in the salt-attracting case of delayed mixing as the Ranz method does not apply.
 

In this article, we study the joint mixing of colloids and salt released together by a two-dimensional (2-D) velocity field $\mathbf{v}(x,y,t)$ with characteristic velocity $V$. In the presence of salt gradients, the colloids do not strictly follow the fluid motions due to electrochemical phenomena but have a velocity $\mathbf{v}_{\mathrm{col}}=\mathbf{v}+ \mathbf{v}_{\mathrm{dp}}$, where $\mathbf{v}_{\mathrm{dp}} = D_\mathrm{dp} \nabla \ln S$ is the diffusiophoretic velocity, $D_\mathrm{dp}$ the diffusiophoretic coefficient and $S$ the total salt concentration (\cite{bib:Anderson1989,bib:Abecassisetal2009}). The concentrations of the salt and colloids, $S$ and $C$, evolve following the set of coupled advection-diffusion equations 
\begin{eqnarray}
&{\displaystyle \partial_t S} + \nabla \cdot  S \mathbf{v} = D_s\, \nabla^2 S,\label{eqs}\\
&{\displaystyle \partial_t C} + \nabla \cdot C (\mathbf{v}+ \mathbf{v}_{\mathrm{dp}}) = D_c\, \nabla^2 C,
\label{eqc}
\end{eqnarray}
where $D_s$, $D_c$ are the diffusion coefficients of each species. 
Note that, although throughout this article we will refer only to diffusiophoresis, the equations are unchanged when dealing with thermophoresis, where particles move under the action of a temperature gradient.

In the following we will consider the special case for which salt and colloids are released {in the same patch with characteristic scale $\ell_0$ at $t=0$}. The time needed to achieve mixing of the colloids, denoted $\tmix$, is governed by the colloids P\'eclet number $Pe_c = V \ell_{0} / D_c$, the salt P\'eclet number $Pe_s = V \ell_{0} / D_s$, and the diffusiophoretic number $D_\mathrm{dp}/D_s$. The case of attenuated mixing will correspond to $D_\mathrm{dp}>0$ (salt-attracting), while enhanced mixing will be modelled by $D_\mathrm{dp}<0$ (salt-repelling). This second situation, also encountered in nature (\cite{bib:Banerjeeetal2016}) and already used in \cite{bib:raynaletal2018},
will allow for a comparison of diffusiophoretic effects in all situations without changing the shape of the profiles. 

The study will be divided into two main sections. In the first section we will address the pedagogical case of mixing of a Gaussian patch, containing salt and colloids, by a linear straining velocity field $\mathbf{v} = (\sigma x, -\sigma y)$ ($\sigma \geq 0$) in the salt-attracting configuration. In this case, the patch of colloids is exponentially stretched in the $x$-direction and compressed towards the modified Batchelor scale $\ell_{c,\mathrm{diff}}=\sqrt{D_c D_s / \sigma (D_\mathrm{dp}+D_s)}$ along the $y$-direction (\cite{bib:raynaletal2018}). In this first example, it is possible to define an effective P\'eclet number for the colloids based on the modified Batchelor scale, $\peff=Pe_c (1+D_\mathrm{dp}/D_s)$, which takes into account the effects of diffusiophoresis. Using this effective P\'eclet number, we find that it is possible to rescale all the values of the mixing time, computed from the analytical solution of \cite{bib:raynaletal2018}, on a single curve derived from the case with no diffusiophoresis.

In the second section, we address the case of chaotic mixing of a sinusoidal patch, with scale $L$, containing salt and colloids by a 2-D time-periodic flow. This second situation is much more complex and requires specific analysis based on the equation for the gradients of concentration to compute an expression for the Batchelor scale. {Following the analysis developed in \cite{bib:raynalgence1997}, which links the mixing time and the time needed for the patch to reach the Batchelor scale, we define the effective P\'eclet number using the modified Bathelor scale in order to rescale the measurements of the mixing time as a function of $\peff$ in all situations. By examining how concentration gradients are generated in the presence of diffusiophoresis,} we derive expressions for the effective P\'eclet number both in the salt-attracting ($\peff = Pe_c D_\mathrm{dp}^2/D_c D_s \gg Pe_c$, attenuated mixing), and in the salt-repelling ($\peff = Pe_c D_c D_s/D_\mathrm{dp}^2 \ll Pe_c$, enhanced mixing) cases. It is shown that these expressions allow collapsing all numerical results for the mixing time in a single curve provided $D_\mathrm{dp}^2/D_c D_s>1$. 


\section{Pure strain}
\label{purestrain}

The case of the joint evolution of Gaussian patches of salt and colloids released at the origin $(0,0)$ in a pure strain flow $\mathbf{v} = (\sigma x, -\sigma y)$ ($\sigma \geq 0$) was solved analytically in \cite{bib:raynaletal2018} so that we only briefly {recall} the steps leading to the solution. 
The P\'eclet numbers are $Pe_s=\sigma \ell_0^2/D_s$ for the salt and $Pe_c=\sigma \ell_0^2/D_c$ for the colloids.
As the salt and colloids concentration profiles are initially Gaussian, they remain Gaussian at all times when deformed by a linear velocity field (\cite{bib:bakuninbook}). Introducing the moments of the salt distribution :
\begin{equation}
\langle x^\alpha y^\beta\rangle_s(t)= \frac{\iint_\infty x^\alpha y^\beta S(x,y,t) dx dy}{\iint_\infty S(x,y,t) dx dy},
\end{equation}
and similar equations for the colloids, the method of moments (\cite{bib:Aris56,bib:birchetal2008}) allows the finding of a closed set of ordinary differential equations (ODEs) for the second-order moments. 

\subsection{Case with no diffusiophoresis}
\label{strainnodif}
\subsubsection{Case of salt}
For an initially round patch $(\langle x^2\rangle_s(0)=\langle y^2\rangle_s(0)=\ell_0^2$, $\langle xy\rangle_s(0))=0$, the solution is
\begin{eqnarray}
\langle x^2\rangle_s(t)&=&\left(\ell_0^2+\frac{D_s}{\sigma}\right)\exp(2\sigma t) -\frac{D_s}{\sigma}\label{eq:x^2sel}\\
\langle y^2\rangle_s(t)&=&\left(\ell_0^2-\frac{D_s}{\sigma}\right)\exp(-2\sigma t) +\frac{D_s}{\sigma}\,.\label{eq:y^2sel}\\
\langle xy\rangle_s(t)&=&0.\label{eq:xysel}
\end{eqnarray}
A large patch such that $\ell_0^2 \gg D_s/\sigma$ (satisfied whenever $Pe_s\gg1$), is exponentially stretched in the dilating direction (since $\langle x^2\rangle_s(t) \approx \ell_0^2 \exp(2\sigma t)$), and exponentially compressed towards the Batchelor scale $\ell_{s} = \sqrt{D_s/\sigma}$ in the compressing direction (following $\langle y^2\rangle_s(t) \approx \ell_0^2 \exp(-2\sigma t) + D_s/\sigma$). 

Neglecting diffusion at short time, it is possible to estimate the time needed for the patch to reach the Batchelor scale under exponential contraction: $\tau_{s} = \ln(\ell_0/\ell_{s})/\sigma$. This expression can be expressed using the P\'eclet number $Pe_s = \ell_0^2/\ell_s^2$ as 
\begin{equation}
\sigma\,\tau_{s}=\frac{1}{2}\ln Pe_s
\label{eq:sigma_tau}
\end{equation}

Of course all the previous reasoning is valid for the mixing of colloids in the absence of diffusiophoresis. By replacing $s$ by $c$ in the expressions, one finds that the P\'eclet number $Pe_c$ is linked to the Batchelor scale $\ell_c$ by the relation 
\begin{equation}
Pe_c=\ell_0^2/\ell_c^2 \,.
\label{eq:Pec_l0^2lc^2}
\end{equation}
This can be also written as 
\begin{equation}
\frac{\ell_{c}}{\ell_0} = \frac{1}{\sqrt{Pe_c}} \, ,
\label{batchcol}
\end{equation} 
which shows that the Batchelor scale of the colloids is smaller than that of the salt as one usually has $Pe_c \gg Pe_s$.

\subsubsection{Mixing time}
It is interesting to note that the time $\tau_{c}$, needed for the patch of colloids to be compressed towards the Batchelor scale, is directly connected to the mixing time, $\tmix$, here defined as the time needed for the concentration, $c(t)$, to decrease by $50\,\%$. Indeed, using equations (\ref{eq:x^2sel}) and (\ref{eq:y^2sel}) one can estimate the concentration of the colloids patch 
\begin{equation}
\frac{c(t)}{c(0)} = \frac{\ell_0^2}{\sqrt{\langle x^2\rangle_c(t)\langle y^2\rangle_c(t)}} \approx\frac{1}{\sqrt{1+1/Pe_c \exp(2\sigma t)}}\,.
\label{eqc(t)}
\end{equation}
At time $t=T_{mix,c}$, $c$ is half of its initial value so that one gets
\begin{equation}
\sigma T_{mix,c}\approx \ln \sqrt{3 Pe_c} = \sigma \tau_{c} + \frac{\ln 3}{2} \, , \label{eq:Tmix_pure_strain_nosalt}
\end{equation}
where $\tau_c$ is the time to reach the Batchelor scale (equation \ref{eq:sigma_tau}).
The two times follow a similar (logarithmic) scaling as functions of the P\'eclet number, and are found to be linked by an affine transformation whose coefficients will depend on the precise definition of the mixing time.

\subsection{Batchelor scale with diffusiophoresis}

When salt and colloids are mixed together, salt is advected by the linear velocity $\mathbf{v} = (\sigma x, -\sigma y)$ while the colloids velocity field is $\mathbf{v}+\mathbf{v}_\mathrm{dp}$. 
As the salt concentration remains Gaussian at all times with $\langle xy \rangle_s=0$, the diffusiophoretic velocity field writes $\mathbf{v}_\mathrm{dp} = (-D_\mathrm{dp} x / \langle x^2 \rangle_s(t),-D_\mathrm{dp} y / \langle y^2 \rangle_s(t))$, which is also a linear flow. The colloids concentration field therefore remains Gaussian at all time, its shape being given by the second-order moments $\langle x^2\rangle_c(t)$, $\langle y^2\rangle_c(t)$, and $\langle xy\rangle_c(t)$. 
In the case of an initially round patch of radius $\ell_0$, one still has $\langle xy\rangle_c(t)=0$ and the equations for $\langle x^2\rangle_c(t)$ and $\langle y^2\rangle_c(t)$ read (\cite{bib:raynaletal2018})
\begin{eqnarray}
\frac{d\langle x^2\rangle_c}{dt}&=&- 2 D_{\mathrm{dp}}\frac{\langle x^2\rangle_c}{\langle x^2\rangle_s}+2\sigma \langle x^2\rangle_c+2D_c
\label{eq:x^2coldef}\\
\frac{d\langle y^2\rangle_c}{dt}&=&- 2 D_{\mathrm{dp}}\frac{\langle y^2\rangle_c}{\langle y^2\rangle_s}-2\sigma \langle y^2\rangle_c+2D_c\:.
\label{eq:y^2coldef}
\end{eqnarray}
This system has a complicated analytical solution given in \cite{bib:raynaletal2018}. 
However the behaviour of its solution may be obtained in the limit of small colloids diffusion coefficient $D_c/D_s \ll 1$ for which salt is mixed in a much shorter time than the colloids so that one has $\langle y^2\rangle_s\simeq\ell_s^2$ for $t > T_{mix,s}$ and
\begin{eqnarray}
\frac{d\langle x^2\rangle_c}{dt}&\simeq& 2\sigma \langle x^2\rangle_c\label{eq:dx2dt}\\
\frac{d\langle y^2\rangle_c}{dt}&\simeq&- 2 \left( \sigma + \frac{D_{\mathrm{dp}}}{\ell_{s}^2} \right) \langle y^2\rangle_c+2D_c \:.\label{eq:dy2dt}
\end{eqnarray}
Equation \ref{eq:dx2dt} shows that the large scale $\langle x^2\rangle_c$ is barely affected by diffusion or diffusiophoresis, a fact that can be easily understood as $x$ is the dilating direction, with exponential stretching by the flow.
In the $y$-direction, using $\ell_{s}^2 = D_s/\sigma$, we see that the colloids concentration field is now compressed towards the modified Batchelor scale 
\begin{equation}
\ell^2_{c,\mathrm{diff}}=\frac{D_c}{\sigma(1+D_\mathrm{dp}/D_s)} \, .
\label{eq:ell_c_modified_strain}
\end{equation}

\subsection{Effective P\'eclet number}
Because we want to quantify mixing through an effective P\'eclet number, and because mixing quantities are closely related to the Batchelor scale, we propose to use equation (\ref{eq:Pec_l0^2lc^2}) and define the effective P\'eclet number $\peff$ using the modified Batchelor scale as
\begin{equation}
\peff=\ell_0^2/\ell_{c,\mathrm{diff}}^2 \,.
\end{equation}
Using equation (\ref{eq:ell_c_modified_strain}), its expression reads 
\begin{equation}
\peff=Pe_c \left(1 + \frac{D_\mathrm{dp}}{D_s} \right).
\label{eq:peff_strain}
\end{equation}

\begin{figure}
\includegraphics{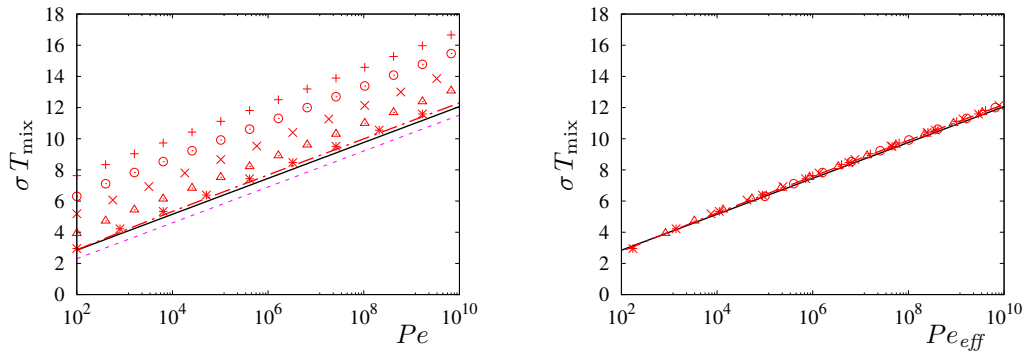}
\caption{
$\sigma\,T_\mathrm{mix}$ for different $D_\mathrm{dp}>0$, $D_s$ and $\sigma$, as a function of the P\'eclet number $Pe$ (left), or of the effective P\'eclet number $Pe_\mathit{eff}$ (right). Black solid line: no salt; \textcolor{red}{--\,$\cdot$\,--} $D_s=1360\,\mu\mathrm{m^2s^{-1}}$, $D_\mathrm{dp}=290\,\mu\mathrm{m^2\, s^{-1}}$;  \textcolor{red}{$\ast$}: $D_s=1360\,\mu\mathrm{m^2s^{-1}}$, $D_\mathrm{dp}=1000\,\mu\mathrm{m^2\, s^{-1}}$;\textcolor{red}{$\triangle$}: $D_s=1360\,\mu\mathrm{m^2s^{-1}}$, $D_\mathrm{dp}=10^4\,\mu\mathrm{m^2\, s^{-1}}$; \textcolor{red}{$\times$}: $D_s=1360\,\mu\mathrm{m^2s^{-1}}$, $D_\mathrm{dp}=10^5\,\mu\mathrm{m^2\, s^{-1}}$; \textcolor{red}{$\circ$}: $D_s=10\,\mu\mathrm{m^2s^{-1}}$, $D_\mathrm{dp}=10^4\,\mu\mathrm{m^2\, s^{-1}}$; \textcolor{red}{$+$}: $D_s=10\,\mu\mathrm{m^2s^{-1}}$, $D_\mathrm{dp}=10^5\,\mu\mathrm{m^2\, s^{-1}}$
\textcolor{magenta}{- - -}: time needed to reach the Batchelor scale; note that this latter follows the same scaling as the mixing time. 
}
\label{fig:mixing_time_Pe_Peeff_salt}
\end{figure}

We have tested this prediction by computing the mixing time of the colloids, defined as the time needed for $c(t)/c(0) = \ell_0^2/\sqrt{\langle x^2\rangle_c(t)\langle y^2\rangle_c(t)}$ to decrease by $50\,\%$. The evolution of $T_{mix,c}$ is displayed in figure \ref{fig:mixing_time_Pe_Peeff_salt} as a function of $Pe_c=\sigma \ell_0^2/D_c$ (left), and as a function of the effective P\'eclet number $\peff=Pe_c (1+D_\mathrm{dp}/D_s)$ (right), for a wide range of parameters. In this case, the mixing time was computed without approximations by using the complicated analytical solution given in \cite{bib:raynaletal2018} where $D_c$, $D_s$ and $D_\mathrm{dp}>0$ were varied over several orders of magnitude. It is remarkable to observe the almost perfect rescaling of $T_{mix,c}$ as a function of $\peff$ spanning over height decades. 

In the case of non-Gaussian patches, because patches tend to relax towards a Gaussian shape after a transient under the combined action of diffusion and stretching by the flow (\cite{bib:villermaux2019}), the scaling given in equation \ref{eq:peff_strain} should remain the same.

Note that in the case of a simple shear, also studied in \cite{bib:raynaletal2018}, the concentration field does not reach a state with a constant Batchelor scale at large time, so that it is not possible to derive an effective P\'eclet number for any linear flow with this method.

\section{Chaotic advection}
In the case of chaotic advection, we also expect the large scale of the colloids patch, governed by exponential stretching by the flow, to be sensitive to neither diffusion nor diffusiophoresis; 
therefore, like in the pedagogical analytical example presented above, diffusiophoretic effects should be visible only when considering the Batchelor scale of the flow.

\subsection{Estimation of the Batchelor scale without diffusiophoresis}

In order to derive the Batchelor scale, it is useful to investigate how small scales are produced which can be done by looking at the equation for the scalar gradient $\mathbf{G}=\nabla C$. In the absence of diffusiophoresis it reads
\begin{equation}
\underbrace{\frac{1}{2} D_t G^2}_{(0)}= \underbrace{D_c\; G_i\partial_j^2 G_i}_{(a)}\underbrace{-G_iG_j\partial_i v_j}_{(b)}\, ,
\label{scalgrad}
\end{equation}
where we have used the convention of summation over repeated indices, and $D_t=\partial_t + v_k \partial_k $ stands for the material derivative. 
In the case of chaotic advection by a large-scale velocity field, small scales are produced by stretching (term $(b)$) until they become so small that the Batchelor scale is reached so that a balance between dissipation and diffusion takes place. 
When mixing is efficient enough, which is the case of global chaos, a quasi-static situation takes place where the left-hand side of (\ref{scalgrad}) is negligible compared to the two other terms so that the balance reads (\cite{bib:raynalgence1997}):
\begin{equation}
\underbrace{D_c\; G_i\partial_j^2 G_i}_{(a)} \approx \underbrace{G_iG_j\partial_i v_j}_{(-b)}\, 
\end{equation}
Given the characteristic length scale $L=\ell_0$ and velocity scale $V$ of the velocity field to get an order of magnitude of the stretching rate $\partial_i v_j  \sim V/L$, one can use this last equation to derive the Batchelor scale for both species:
\begin{eqnarray}
\ell_{c}&\sim& \frac{L}{\sqrt{Pe_c}} \hspace{1cm} \text{(colloids, no diffusiophoresis}),\label{eq:small_scale_no_diffusio} \\
\ell_{s}&\sim& \frac{L}{\sqrt{Pe_s}} \hspace{1cm} \text{(salt}),\label{eq:small_scale_salt}
\end{eqnarray}
with $Pe_c=VL/D_c$ and $Pe_s=VL/D_s$. Using these relations, the time needed to mix a patch of scalar of size $L$ will follow the same scaling relation as the time $\tau_B = \ln Pe / \Lambda$, needed to reach the Batchelor scale, where $\Lambda \propto V/L$ is the most negative Lyapunov exponent of the flow (\cite{bib:raynalgence1997,bib:bakuninbook}). In the case of joint mixing with diffusiophoresis, {the Batchelor scale ($\ell_{c,\mathrm{diff}}$) will have a different expression so that the relation (\ref{eq:small_scale_no_diffusio}) is not expected to hold anymore. However} we shall follow the path of the previous section to derive an effective P\'eclet number based on the modified Batchelor scale through the relation 
\begin{equation}
\ell_{c,\mathrm{diff}} \sim L/\sqrt{\peff} \label{eq:small_scale_diffusio}
\end{equation}
so as to rescale all measurements of the mixing time as a function of $\peff$.

\subsection{Batchelor scale in the case of diffusiophoresis}
In absence of diffusiophoresis, we found the Batchelor scale to be $\ell_{c}=L/\sqrt{Pe_c} = \sqrt{D_c/\sigma_v}$ with $\sigma_v=V/L$. This expression being the same as the one found for the pure strain flow of section (\ref{strainnodif}), it may be anticipated that the effective P\'eclet number for the mixing of colloids is given by the relation $\peff = Pe_c (1+D_\mathrm{dp}/D_s)$. However as we shall see in the following, the simple result of section (\ref{strainnodif}) will not hold in the case of chaotic advection. In order to derive a correct expression for the Batchelor scale with diffusiophoresis, we must again look carefully at the equation for $\nabla C$ in the presence of the velocity drift $\vdp=D_\mathrm{dp} \nabla \ln S$.
Taking the gradient of equation (\ref{eqc}), we obtain
\begin{equation}
\partial_t G_i+v_j\partial_j G_i+v_{{\rm dp}\,j}\partial_j G_i+G_i\partial_j v_{{\rm dp}\,j}+G_j\partial_i v_j+G_j\partial_i v_{{\rm dp}\,j}+C\partial_i\partial_j v_{{\rm dp}\,j}=D_c\; \partial_j^2 G_i\,,
\end{equation}
which we shall multiply by $G_i$ and sum over all components to get
\begin{equation}
\underbrace{\frac{1}{2}D_t G^2}_{(0)}= \underbrace{D_c\; G_i\partial_j^2 G_i}_{(a)}\underbrace{-G_iG_j\partial_i v_j}_{(b)}\underbrace{-v_{{\rm dp}\,j}G_i\partial_j G_i}_{(c)}\underbrace{-G^2\partial_j v_{{\rm dp}\,j}}_{(d)}\underbrace{-G_iG_j\partial_i v_{{\rm dp}\,j}}_{(e)}\underbrace{-CG_i\partial_i\partial_j v_{{\rm dp}\,j}}_{(f)}\,.
\label{eq:gradients_modulus}
\end{equation}

As opposed to equation (\ref{scalgrad}),  equation (\ref{eq:gradients_modulus}) contains four additional terms involving the diffusiophoretic drift $\vdp=D_\mathrm{dp} \ln S$. In order to get an order of magnitude of each term, one needs to estimate the magnitude of $\vdp$. This can be done in the case $D_c \ll D_s$ for which the salt patch is compressed towards $\ell_s$ long before diffusion affects the colloids concentration field. The magnitude of diffusiophoretic drift is then
\begin{equation}
V_{\rm dp}\sim \frac{D_{\rm dp}}{\ell_s} = \frac{D_{\rm dp}}{L}\sqrt{Pe_s} \, ,
\end{equation}
in agreement with the numerical simulations of \cite{bib:Volketal2014}. 

Estimating the different terms of equation (\ref{eq:gradients_modulus}) in the quasistatic regime, when the colloids patch has been compressed towards its Batchelor scale $\ell_{c,\mathrm{diff}}$ so that $G\sim C/\ell_{c,\mathrm{diff}}$, we get
\begin{eqnarray}
(a) &\sim& \frac{G^2 D_c}{\ell_{c,\mathrm{diff}}^2} = \frac{G^2D_c}{L^2} Pe_\textit{eff}\\
(b) &\sim& \frac{G^2 V}{L} = \frac{G^2D_c}{L^2} Pe_c\\
(c) &\sim& \frac{G^2}{\ell_{c,\mathrm{diff}}}\frac{D_\mathrm{dp}}{\ell_s} = \frac{G^2D_c}{L^2} \frac{D_\mathrm{dp}}{\sqrt{D_c D_s}}\sqrt{Pe_c\, Pe_\textit{eff}}\\
(d) &\sim& \frac{G^2 D_{\rm dp}}{\ell_s^2} \ll (c) \\
(e) &\sim& \frac{G^2 D_{\rm dp}}{\ell_s^2} \ll (c) \\
(f) &\sim& G^2 \ell_{c,\mathrm{diff}} \frac{D_{\rm dp}}{\ell_s^3} \ll (c)
\end{eqnarray}
As one has $\ell_{c,\mathrm{diff}} \ll \ell_s$, the last three terms are always much smaller than $(c)$ so that they will be neglected in the analysis.

\subsection{Effective P\'eclet number in the salt-attracting case ($D_\mathrm{dp}>0$)}

This configuration, which was not addressed in the case of chaotic advection, is similar to the case of pure deformation addressed in section (\ref{purestrain}). As colloids and salt are released together, the velocity drift is opposed to molecular diffusion so that the mixing time of colloids increases compared to the case with no diffusiophoresis, leading to 
\begin{equation}
Pe_\mathit{eff}^\mathit{attract}\gg Pe_c \, . \label{eq:peff_gg_pec}
\end{equation}
In the quasistatic regime, where production and dissipation balance, the dominant terms are then $(a)$ and $(c)$ so that filaments are produced by diffusiophoresis and dissipated by diffusion, as was the case in the analytical example. Equating $(a)$ and $(c)$, we get
\begin{equation}
Pe_\mathit{eff}^\mathit{attract}\sim\frac{D_\mathrm{dp}^2}{D_c D_s}Pe.
\label{eq:Pe_eff_salt_in}
\end{equation}
Note that, because of hypothesis (\ref{eq:peff_gg_pec}) the analysis can only be true if one has ${D_\mathrm{dp}^2} \gg {D_c\,D_s}$, which is satisfied in experiments as $D_\mathrm{dp} \sim D_s$ and $D_c \ll D_s$ (\cite{bib:Abecassisetal2009,bib:Deseigneetal2014}).

The effective P\'eclet number found here is completely different from that of equation \ref{eq:peff_strain} (linear strain). 
This is not surprising: whilst both flows display exponential stretching, in the chaotic advection case, as explained, a quasi-stationary state is reached where gradients of concentration are created by the flow and dissipated by diffusion at the same rate. 
This is completely different in the case of the simple strain: because this linear flow is barely mixing, when the Batchelor scale is reached, all terms in the equation of gradients \ref{scalgrad} decay at the same exponential rate, so that the left-hand-side (term (0)) is far from negligible compared to the two others.

\subsection{Effective P\'eclet number in the salt-repelling case ($D_\mathrm{dp}<0$)}
This second configuration was first addressed in (\cite{bib:Deseigneetal2014}) where the authors proposed an expression $Pe_\mathit{eff}^\mathit{repell} \propto VL/\deff$, with $\deff \propto D_\mathrm{dp}^2/D_s$, based on the Ranz model of mixing (\cite{bib:ranz1979}). 
In this configuration, diffusiophoresis acts as an enhanced diffusion so that we will assume 
\begin{equation}
Pe_\mathit{eff}^\mathit{repell}\ll Pe\, , 
\label{eq:peff_ll_pec}
\end{equation}
and the quasistatic equilibrium is now given by $(b) \sim (c)$, with gradients produced by the flow and smoothed by diffusiophoresis. 
Under this assumption we obtain the same result as proposed in (\cite{bib:Deseigneetal2014}):
\begin{equation}
Pe_\mathit{eff}^\mathit{repell}\sim\frac{D_c D_s}{D_\mathrm{dp}^2}\,Pe_c\,.
\label{Pe_eff_salt_out}
\end{equation}
which, using hypothesis (\ref{eq:peff_ll_pec}), is found to hold under the same hypothesis ${D_\mathrm{dp}^2} \gg {D_c\,D_s}$ as in the salt-attracting case.

\subsection{Comparison with numerical simulations} 
We have tested the two scaling relations in numerical simulations of chaotic mixing by a sine flow with random phase, which ensures that global chaos is achieved (\cite{bib:pierrehumbert1994,bib:pierrehumbert2000}) so that the mixing time is expected to scale linearly with the logarithm of the P\'eclet number. The flow is a modified version of the one used in (\cite{bib:Volketal2014}): it is composed of two sub-cycles of duration $T/2$ for which the velocity field is $\v(\mathbf{r},t)=(f(t) \sin(y+\phi_n),0)$ for $nT\leq t < (n+1/2)T$ and  $\v(\mathbf{r},t)=(0,f(t)\sin(x+\psi_n))$ for $(n+1/2) T\leq t < (n+1)T$, where $(\phi_n,\psi_n)_{n \geq 1}\in [0, 2 \pi]^2$ are sequences of random numbers and $f(t)=2 \sin^2(2\pi t/T)$. Because $\int_0^{T/2} f(t) \mathrm{d}t =1$, this flow corresponds to the same iterated map as the one with $f(t)=1$ without temporal discontinuities.

The equations were solved in a square periodic domain of length $L=2 \pi$, with $T=1.6 \, \pi$ and identical {concentration profiles $C(x,y,t=0)=S(x,y,t=0)=1+\sin x$}, with the same code used in (\cite{bib:Volketal2014}). In order to test the scaling relations, we performed a set of simulations while keeping the same sequence of random phases, varying $D_c \in [2 \,10^{-5}, 8 \,10^{-4}]$, $D_s = (0.25 \, 10^{-2}, \, 0.5 \, 10^{-2}, \, 10^{-2})$, and $D_\mathrm{dp} =  (\pm \, 10^{-3}, \, \pm 2 \, 10^{-3}, \pm 4 \, 10^{-3})$. This corresponds to P\'eclet numbers $Pe_c \in [7.8 \,10^3 - 3\, 10^5]$ for the colloids, $Pe_s\in [600-2500]$ for the salt. 

\begin{figure}
\includegraphics{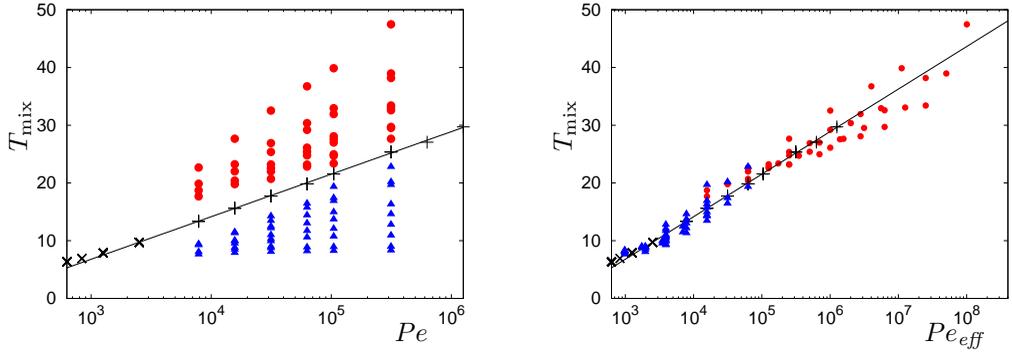}
	\caption{
{Left: mixing time of the colloids, $T_\mathrm{mix,c}$, as a function of the P\'eclet number for all numerical simulations with $D_\mathrm{dp}^2/(D_cDs)>1$; the evolution of the flow is kept identical, varying $D_c$, $D_\mathrm{dp}$, and $D_s$. Right: $T_\mathrm{mix,c}$ as a function of the effective P\'eclet number. $+$: mixing of colloids without diffusiophoresis ($D_\mathrm{dp}=0$);  $\times$: mixing of salt ; \textcolor{red}{$\bullet$}: ``salt-attracting" case ($D_\mathrm{dp}>0$); \textcolor{blue}{$\blacktriangle$}: ``salt-repelling" case ($D_\mathrm{dp}<0$). The solid line is a curve of expression $T_\mathrm{mix,c}=3.2 \ln(Pe/120)$.} \label{fig:Tmix_Pe_Pe_eff}
}
\end{figure}
The evolution of the mixing time of the colloids, $T_{mix,c}$, as a function of $Pe_c$ is displayed in figure (\ref{fig:Tmix_Pe_Pe_eff}, left) for all simulations with {$D_\mathrm{dp}^2/D_cD_s > 1$}, and different symbols for the salt-attracting case (circles) and the salt-reppelling case (triangles). The solid line is a fit for the non-diffusiophoretic cases (salt or colloids without salt), defined as 
\begin{equation} 
T_{mix}=3.2\, \log(Pe/120)
\label{tmixcol}
\end{equation}
for this particular flow. 

As shown in figure (\ref{fig:Tmix_Pe_Pe_eff}, right), all values of the mixing time are found to follow well the curve obtained without diffusiophoresis when they are plotted as a function of the effective P\'eclet numbers $Pe_\mathit{eff}^\mathit{attract}=\frac{D_\mathrm{dp}^2}{D_c D_s}\,Pe$ for ``salt-attracting" case, and $Pe_\mathit{eff}^\mathit{repell}=\frac{D_c D_s}{D_\mathrm{dp}^2}\,Pe$ for the ``salt-repelling" case; 
note also that the effective P\'eclet number varies here over five orders of magnitude. 

\begin{figure}
\begin{center}
\includegraphics[width=65mm]{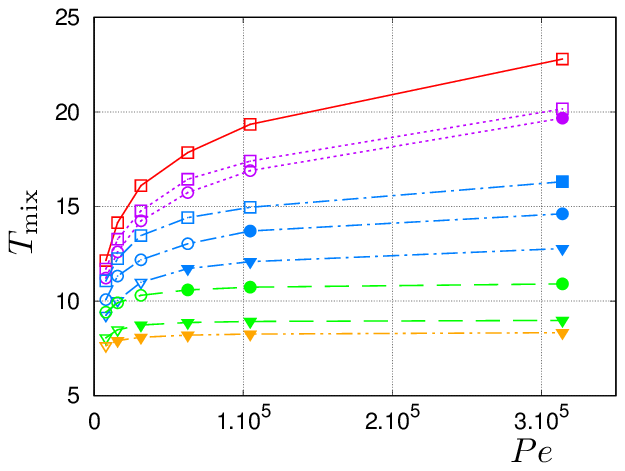}
\end{center}
	\caption{
Mixing time $T_\mathrm{mix}$ as a function of the P\'eclet number for different values of $D_\mathrm{dp}$ and $D_s$ in the salt-repelling case ($D_\mathrm{dp}<0$); $\square$: $D_\mathrm{dp}=10^{-3}$; $\circ$: $D_\mathrm{dp}=2.\,10^{-3}$; $\triangledown$:   $D_\mathrm{dp}=4.\,10^{-3}$; 
\textcolor{red}{---}: $D_\mathrm{dp}/D_s=0.1$; \textcolor{purple}{$\cdot\cdot\cdot\cdot$}: $D_\mathrm{dp}/D_s=0.2$; \textcolor{blue}{$- \cdot -$}: $D_\mathrm{dp}/D_s=0.4$; \textcolor{green}{$- -$}: $D_\mathrm{dp}/D_s=0.8$; \textcolor{orange}{--$\,\cdot\, \cdot\,$--}: $D_\mathrm{dp}/D_s=1.6$
The full symbols are those for which $D_\mathrm{dp}^2/(D_cDs)\ge10$; open symbols: $D_\mathrm{dp}^2/(D_cDs)<10$. 
\label{fig:Tmix_Pe_eff_salt_repelling_plateau}
}
\end{figure}

The validity of our analysis can be further checked by focusing on the salt-repelling case. In that case the relation (\ref{Pe_eff_salt_out}) is indeed equivalent to having an effective diffusivity
\begin{equation}
D_\mathit{eff}^\mathit{repell}\sim\frac{D_\mathrm{dp}^2}{D_s}\ ,
\label{eq:Deff}
\end{equation}
or an effective P\'eclet number independent of $D_c$, such that the mixing time becomes independent of $Pe_c$. Figure \ref{fig:Tmix_Pe_eff_salt_repelling_plateau} displays $T_{mix,c}$, as a function of $Pe_c$ in the salt-repelling case ($D_\mathrm{dp}<0$), linking the data points corresponding to the same values of $D_\mathrm{dp}$ and $D_s$. As predicted, the different curves seem to exhibit a plateau when $Pe_c$ is large enough so that ${D_\mathrm{dp}^2}/{D_c D_s} \geq 10$ (filled symbol). 

\section{Summary and conclusion}

We have studied the joint mixing of colloids and salt in a stagnation point and in a globally chaotic flow, and investigated how the mixing time is modified by varying the diffusivities of the colloids and salt diffusivities, $D_c$ and $D_s$, and the diffusiophoretic coefficient $D_\mathrm{dp}$.

In the case of the dispersion of Gaussian patches in a pure deformation flow, diffusiophoresis led to a modification of the Batchelor scale $\ell_{c,\mathrm{diff}} = \sqrt{D_cD_s/\sigma (D_\mathrm{dp} + D_s)}$ which was used to derive an effective P\'eclet number $\peff$ through the relation $\ell_{c,\mathrm{diff}}/\ell_0 = 1 / \sqrt{\peff}$. 
It was thus possible to achieve a remarkable rescaling of the mixing time as a function of $\peff$, on the same curve as the one used in the absence of diffusiophoresis, as obtained from the analytical solution of \cite{bib:raynaletal2018}.

The case of chaotic advection being far more complex than the one of a linear velocity field, we used the equation for the scalar gradients to derive expressions for the Batchelor scale, by balancing production and dissipation of scalar gradients in the presence of a diffusiophoretic drift with magnitude 
$V_{\rm dp}\sim {D_{\rm dp}}/{\ell_s}$.
The approach allowed us to define an effective P\'eclet number $Pe_\mathit{eff}^\mathit{repell}\sim Pe_c \,{D_c D_s}/{D_\mathrm{dp}^2}$ in the salt-repelling ($D_\mathrm{dp}<0$) case, which is the same expression as the one proposed by \cite{bib:Deseigneetal2014} based on the approach of Ranz. However our analysis was not limited to that configuration and we derived an expression for the salt-attracting case $Pe_\mathit{eff}^\mathit{attract}\sim Pe_c \,{D_\mathrm{dp}^2}/{D_c D_s}$, both expressions being valid under the same assumption ${D_\mathrm{dp}^2}/{D_c D_s} \gg 1$.
The prediction was tested using numerical simulations of chaotic advection, which allowed us to compute the mixing time for a large range of parameters $D_c$ , $D_s$ and $D_\mathrm{dp}$, and we observed a very good rescaling of the results provided ${D_\mathrm{dp}^2}/{D_c D_s} > 1$. It confirmed the observation of \cite{bib:Deseigneetal2014} who found that the mixing time was almost independent of $D_c$ (${D_\mathrm{dp}^2}/{D_c D_s} \simeq 30$ in their experiments).

In this second case, both patches of salt and of colloids were injected at the scale of the flow and mixed due to Lagrangian chaos so that the effective P\'eclet number was obtained using an argument based on the Batchelor scale. Another situation, which deserves further attention, corresponds to the dispersion of patches of scalars whose spatial variations are much larger than the flow scale. 
In this large-scale dispersion problem, effective diffusivities would typically be obtained using an homogenization method (\cite{frishbook95,Biferale_etal95}) with imposed mean gradients of salt and colloids. It would then be interesting to investigate if the various methods lead to similar scalings.

Another interesting aspect concerns the mixing of salt and colloids in a turbulent flow. This is of course the most challenging as diffusiophoresis leads to clustering of colloids at very small scales due to the large value of the Schmidt number, $Sc=\nu/D_s$ (with $\nu$ the kinematic viscosity), for any salt dissolved in water (\cite{Schmidt2016,bib:Vishwanathetal2017}). 
If the two scalings derived in the present article would not be correct in the turbulent case, the present method is not limited to laminar flows so that it could be possible to derive an effective P\'eclet number for the turbulent case following \cite{bib:raynalgence1997}.\\

\noindent\textbf{Acknowledgements}\\
The authors benefited from fruitful discussions with M. Bourgoin, C. Cottin-Bizonne, and C. Ybert. This work was supported by the French research programs ANR-16-CE30-0028, and IDEXLYON of the University of Lyon in the framework of the French program ``Programme Investissements d'Avenir" (ANR-16-IDEX-0005).

\bibliography{Compressibility}
\bibliographystyle{jfm}

\end{document}